\documentclass[doublecol]{epl2} 
\usepackage{hyperref}
\usepackage{subfigure}
\usepackage{subcaption}
\usepackage[normalem]{ulem}

\usepackage[numbers]{natbib}
\title{Two-Temperature Induced Phase Separation: Non-equilibrium Phase Behavior, Ordering, and Kinetics}
\shorttitle{Two-Temperature Induced Phase Separation: Non-equilibrium Phase Behavior, Ordering, and Kinetics} 

\author{
Nayana Venkatareddy\inst{1,3}$^{*}$
\and Jaydeep Mandal\inst{1}$^{*}$
\and Jayeeta Chattopadhyay\inst{2}$^{*}$
\and Prabal K. Maiti\inst{1}\thanks{Corresponding author: maiti@iisc.ac.in}
}

\institute{
\inst{1} Department of Physics, Indian Institute of Science,
C. V. Raman Ave, Bengaluru 560012, India\\
\inst{2} Physikalisches Institut, Albert-Ludwigs- Universität Freiburg, D-79104 Freiburg, Germany\\
\inst{3} Current address: Department of Physics, Technion- Israel Institute of Technology, Haifa 3200003, Israel.
}
\abstract{Two-temperature induced phase separation (2-TIPS) has emerged as a generic non-equilibrium mechanism in scalar active systems with heterogeneous activity, where particles coupled to different thermal reservoirs spontaneously demix into dense cold and dilute hot phases. Unlike equilibrium phase separation or motility-induced phase separation (MIPS), 2-TIPS is driven solely by unequal energy injection and the resulting heat flux between particle species. This review summarizes recent advances in 2-TIPS across diverse soft-matter systems, highlighting both its universal non-equilibrium mechanisms and the emergent ordered phases arising from particle shape anisotropy, chirality, confinement, and topology. We further discuss density-dependent phase-separation kinetics and coarse-grained descriptions linking microscopic dynamics to macroscopic behavior, and outline key directions for future research.
}

\begin{document}

\maketitle
\noindent
$^{*}$ These authors contributed equally to this work.

\section{Introduction}
Phase separation \cite{bray1994theory,Onuki_2002,Cates_Tjhung_2018} is one of the most ubiquitous self-organization phenomena in nature, occurring across a vast range of spatial and temporal scales. Examples range from gas–liquid coexistence \cite{10.1063/1.4720089} in simple fluids and demixing in multi-component liquid or solid mixtures \cite{puri2009kinetics} to the formation of intracellular condensates, membrane-less organelles, vesicular compartments, and cell sorting in biological tissues \cite{annurev:/content/journals/10.1146/annurev-cellbio-100913-013325,Yoshizawa2020,Bergeron-Sandoval2016}. Under equilibrium conditions, phase separation is generally driven by differences in intermolecular interactions, particle shapes, or entropic effects arising from excluded-volume constraints \cite{PhysRevE.54.605, poon_1994_phase,10.1063/1.4801333}. The resulting coexistence of distinct phases can be understood within the framework of equilibrium statistical mechanics \cite{pathria2017statistical} through the minimization of thermodynamic free energy. However, many natural and synthetic systems continuously consume energy and operate far from thermodynamic equilibrium, rendering the equilibrium framework insufficient to describe their collective behavior. Active matter \cite{annurev:/content/journals/10.1146/annurev-conmatphys-070909-104101,prost2015active}
constitutes a broad class of such non-equilibrium systems, whose constituent units continuously convert stored or ambient energy into mechanical motion or work at the microscopic scale. These systems are also found across a range of length scales, from flocks of birds, schools of fish, and bacterial colonies \cite{VICSEK201271,toner2005hydrodynamics} to intracellular structures such as the cytoskeleton  \cite{annurev:/content/journals/10.1146/annurev-conmatphys-031218-013231} and chromatin \cite{ganai2014chromosome}. Advances in experimental techniques have also enabled the realization of synthetic active systems, including self-propelled colloids and Janus particles, with promising applications in targeted drug delivery, microscale transport, and environmental remediation \cite{Wang2019,PhysRevX.9.041032,GHOSH2020100836}. Active matter systems also exhibit fascinating collective phenomena with no equilibrium counterpart, including flocking \cite{VICSEK201271}, swarming, active turbulence \cite{doi:10.1073/pnas.1202032109}, and motility-induced phase separation (MIPS) \cite{cates2015motility}.  

One of the most prominent examples of non-equilibrium self-organization is motility-induced phase separation (MIPS) \cite{cates2015motility,redner2013structure,fily2012athermal}, in which self-propelled particles interacting solely through repulsive forces spontaneously separate into dense and dilute phases. Persistent self-propulsion causes particles to accumulate in regions where their motion is hindered by neighboring particles, creating a positive feedback between local density and motility suppression. While MIPS has emerged as the paradigmatic example of activity-induced phase separation, it relies on vectorial activity arising from directed self-propulsion. Many non-equilibrium systems, however, lack intrinsic directional motion and are more naturally described by scalar activity. For example, gene-rich chromatin regions \cite{ganai2014chromosome} experience enhanced stochastic forces due to ATP-dependent transcriptional activity, catalytic enzymes in the presence of substrates \cite{PhysRevLett.123.128101} and passive tracers in active media \cite{PhysRevLett.84.3017,CHAKI2019121574,CHAKI2018302} exhibit enhanced diffusion, while mixtures of active and passive colloids \cite{PhysRevE.101.022606} possess markedly different diffusivities \cite{weber2016binary}. In such systems, activity manifests through differences in fluctuation strength, transport coefficients, or effective temperatures. A related realization of scalar activity arises in responsive colloids, where internal particle degrees of freedom are coupled to a thermal reservoir distinct from that governing translational motion, generating non-equilibrium behavior without self-propulsion \cite{PhysRevE.106.014613}. Consequently, heterogeneous active systems can often be modeled as mixtures of particles maintained at different effective temperatures or diffusivities \cite{weber2016binary,smrek2017small,chari2019scalar}. Similar multi-temperature descriptions also arise in structural and spin glasses \cite{PhysRevE.55.3898}, as well as in systems with non-reciprocal interactions, dusty plasmas, granular mixtures, and chemically driven colloidal suspensions \cite{PhysRevX.5.011035,PhysRevLett.92.164301,PhysRevLett.105.088304}. Remarkably, even in the absence of self-propulsion or alignment interactions, such systems undergo spontaneous segregation, giving rise to two-temperature induced phase separation (2-TIPS) \cite{weber2016binary,chari2019scalar}.

Given its simplicity and broad applicability, 2-TIPS has attracted considerable theoretical \cite{grosberg2018dissipation,grosberg2015nonequilibrium,PhysRevE.101.022120} and computational \cite{smrek2017small, chari2019scalar,smrek2018interfacial,PhysRevE.107.034607, chattopadhyay2021heating,PhysRevE.107.024701,D3SM00796K,chattopadhyay2024stability,10.1063/1.4720089,1c8b-hmxv,venkatareddy2025phaseseparationkinetics2tips,mccarthy2024demixing} interest over the past decade. The earliest evidence for activity-induced segregation in scalar systems was provided by Weber and co-workers \cite{weber2016binary}, who demonstrated that mixtures of particles differing only in their diffusivities spontaneously demix into dense clusters of low-diffusivity particles surrounded by a dilute phase of highly diffusive particles. Subsequent studies extended this idea to polymeric systems, where Kremer \cite{smrek2017small,smrek2018interfacial} and collaborators showed that activity differences can drive phase segregation in active-passive polymer blends and influence interfacial properties and entropy production.
 On the theoretical front,  Joanny \cite{grosberg2018dissipation,grosberg2015nonequilibrium} and co-workers developed a continuum theory for mixtures of particles coupled to different temperatures, showing that temperature differences alone can drive phase separation even in systems with purely excluded-volume interactions. In the dilute limit, they demonstrated that the non-equilibrium 2-TIPS admits an effective Cahn–Hilliard description, enabling predictions for phase diagrams, interfacial tension, and coarsening dynamics. From a Hamiltonian perspective, Netz \cite{PhysRevE.101.022120} developed a theoretical framework for interacting many-particle systems coupled to multiple heat baths, introducing a free-entropy formalism to describe non-equilibrium stationary states and entropy production. Maiti and co-workers \cite{chari2019scalar,PhysRevE.107.034607, chattopadhyay2021heating,PhysRevE.107.024701,D3SM00796K,chattopadhyay2024stability,10.1063/1.4720089,1c8b-hmxv,venkatareddy2025phaseseparationkinetics2tips}, using extensive molecular dynamics (MD) simulations, introduced the 2-TIPS framework to a wide range of soft matter systems, including liquid crystals and confined geometries, revealing rich structural, dynamical, and ordering phenomena.

In this review, we first introduce the general framework and salient features of 2-TIPS before examining its behavior across a broad range of soft matter systems, including isotropic Lennard-Jones (LJ) fluids \cite{chari2019scalar}, dumbbells \cite{PhysRevE.107.034607}, soft spherocylinders (SRSs) \cite{chattopadhyay2021heating,PhysRevE.107.024701}, and chiral helices \cite{chattopadhyay2024stability}. We discuss how particle geometry, interaction anisotropy, and confinement \cite{D3SM00796K} influence phase behavior and self-organization, and review recent advances in the non-equilibrium kinetics of 2-TIPS, including domain growth and coarsening \cite{1c8b-hmxv,venkatareddy2025phaseseparationkinetics2tips}. Finally, we highlight open questions and future directions for scalar active systems.

\section{General Framework of 2-TIPS}
In 2-TIPS, non-equilibrium properties are studied in binary mixtures composed of two species maintained at different temperatures \cite{chari2019scalar}. In numerical simulations, this is typically achieved by partitioning the system into two subsets of particles, commonly referred to as `hot' and `cold', which are coupled to thermostats at temperatures \(T_h\) and \(T_c\), respectively, with $T_h > T_c$. The particles interact via identical inter- and intra-species interactions, ensuring that the observed phase behavior is driven by temperature disparity rather than interaction differences. Depending on the system under consideration, the temperature disparity can be imposed using Nos\'e--Hoover\cite{10.1063/1.449071}, Berendsen \cite{berendsen1984molecular}, or Langevin thermostats \cite{weber2016binary}. While the choice of thermostat can significantly influence the kinetics, the steady-state behavior remains qualitatively similar across different implementations.

\section{Steady-State Behavior}
The continuous exchange of energy between hot and cold particles drives the system into a non-equilibrium steady state characterized by persistent heat flux, positive entropy production, and spontaneous segregation into dense cold and dilute hot domains. Below, we summarize the salient steady-state features of 2-TIPS that are consistently observed across different soft-matter systems.

\textit{Effective temperatures and activity:} 
The presence of two species maintained at distinct temperatures leads to a sustained heat flux (power transfer) \cite{PhysRevE.107.024701} from hot to cold particles via collisions, even in steady state.  Consequently, the effective temperatures of the two species, estimated from their velocities via the equipartition theorem, deviate from their respective thermostat values \cite{chari2019scalar}. Specifically, the effective temperature of cold particles \(T_c^{eff}\) is higher than the imposed temperature \(T_c\), while the opposite holds true for hot particles, \(T_h^{eff}<T_h\). The deviation of the effective temperatures depends on both the magnitude of power transfer and the coupling constant of the thermostat used. The phase behavior of the mixture is governed by the relative temperature difference between the species, characterized by the activity \(\chi=(T_h^{eff}-T_c^{eff})/T_c^{eff}\), rather than by the absolute values of the thermostat temperatures.

\textit{Extent of phase-separation and critical activity:} For small activity $\chi$, the system remains homogeneous.  The critical activity \(\chi_c\) is defined as the lowest value of \(\chi\), at which phase separation becomes macroscopic. When $\chi>\chi_c$, macroscopic phase separation sets in \cite{chari2019scalar,smrek2017small,chattopadhyay2021heating}, and the degree of segregation increases with activity as depicted in Fig. \ref{fig:1}(a). Additionally, dimensionality plays a crucial role: in three-dimensional (3D) bulk systems \cite{chari2019scalar}, phase separation is enhanced with increasing density, whereas in two-dimensional systems (2D) \cite{D3SM00796K}, the opposite trend is often observed. This reversal arises because reducing the dimensionality changes the interface between the hot and cold phases from a surface in 3D to a line in 2D, making it more difficult for trapped hot particles to escape dense cold clusters as density increases, and thereby reducing the overall extent of phase separation. 

The critical activity varies across different soft matter systems \cite{chari2019scalar,PhysRevE.107.034607,chattopadhyay2021heating} and exhibits a nontrivial density dependence \cite{chattopadhyay2021heating}. In the liquid regime, \(\chi_c\) decreases with increasing density because more frequent collisions promote energy transfer from hot to cold particles, facilitating phase separation. At sufficiently high densities, however, the trend reverses and \(\chi_c\) increases again because hot particles become trapped within the ordered cold domains and the slow relaxation of these phases also hinders phase separation.

\textit{Coexistence at different densities:}
A distinctive feature of 2-TIPS is the coexistence of phase-separated hot and cold particles at different densities \cite{chari2019scalar}. The cold particles aggregate into dense regions with densities exceeding the system average, while the hot particles expand into a dilute density. The dense cold phase often exhibits enhanced structural ordering relative to the homogeneous state, whereas the hot phase remains disordered. This density contrast gives rise to a rich variety of emergent structures, particularly in systems with anisotropic particle shapes \cite{PhysRevE.107.024701}, as discussed in later sections.

\textit{Mechanical equilibrium and pressure balance}: Although the phase-separated domains coexist at different densities and temperatures, the pressure (particularly the normal pressure) remains nearly constant across the interface in mechanical equilibrium \cite{chattopadhyay2021heating,PhysRevE.107.034607}. This is achieved through a balance between the high kinetic pressure in the hot, dilute regions and the high virial pressure in the dense cold regions. So in 2-TIPS, any pressure differences arising from temperature differences vanish upon phase separation, yielding coexisting dense cold and dilute hot domains in mechanical equilibrium. 

\textit{Heat transfer and entropy production}:
While mechanical equilibrium is established across the phase-separated domains, the system remains intrinsically out of equilibrium due to the continuous transfer of energy from the hot to the cold particles. This persistent heat flux is finite not only at the interface, where hot and cold particles interact directly, but also throughout the bulk phases \cite{PhysRevE.107.024701} resulting in heterogeneous activity throughout the system. The direction of the heat current is from the bulk hot region toward the bulk cold region. Consequently, the system exhibits positive entropy production \cite{chari2019scalar, PhysRevE.107.034607} and broken time-reversal symmetry, hallmarks of a non-equilibrium steady state. 

\subsection{Steady states in 2-TIPS across different soft matter systems} 
Although 2-TIPS exhibits several universal features, important aspects of its behavior depend on the specific soft matter system.

 \textbf{a.}    LJ systems \cite{chari2019scalar} provide the prototypical model for 2-TIPS in particles with isotropic interactions. In both 2D and 3D, phase separation produces dense crystalline domains of cold particles. In 3D these domains predominantly exhibit hexagonal close-packed (HCP) and face-centered cubic (FCC) order, while in 2D they form hexagonally ordered crystals, consistent with the equilibrium phases of dense LJ systems.

\textbf{b.} Since shape anisotropy is ubiquitous in soft matter, 2-TIPS has also been studied in dumbbells \cite{PhysRevE.107.034607}, the simplest model of anisotropic particles consisting of two bonded monomers. Compared to monomers, dumbbells require a higher critical activity $\chi_c$ for phase separation at all densities, demonstrating that shape anisotropy provides an additional control parameter for non-equilibrium phase behavior. Symmetric dumbbells form crystalline cold domains with predominantly HCP and FCC ordering, whereas asymmetric dumbbells exhibit both crystalline and non-crystalline cold domains depending on their shape asymmetry.

\textbf{c.} Beyond dumbbells, the effects of two-temperature activity become even richer in systems of anisotropic rod-like particles as demonstrated in Fig. \ref{fig:1} (b). MD simulations of soft repulsive spherocylinders (SRSs) \cite{chattopadhyay2021heating} have shown that 2-TIPS not only induces phase separation but also drives activity-induced liquid-crystalline (LC) ordering in the cold domains. The critical activity \(\chi_c\) for phase separation was found to lie within a relatively small range (compared to monomeric systems), 
    over a wide range of densities spanning isotropic (I), nematic (N), smectic (Sm), and crystalline (K) equilibrium phases. Starting from a homogeneous isotropic phase at a fixed temperature, the system spontaneously separates into hot and cold regions, where the cold particles develop LC order while the hot particles remain comparatively dilute and isotropic. More generally, starting from isotropic, nematic, or smectic equilibrium states, activity drives the cold particles toward progressively more ordered phases, including nematic, smectic, and crystalline structures, whereas the hot particles evolve toward less ordered states compared to the initial equilibrium configuration. As a consequence, the isotropic--nematic phase boundary shifts toward lower densities for the cold particles and toward higher densities for the hot particles relative to its equilibrium location. Remarkably, two-temperature activity can even induce liquid-crystal phases that are inaccessible in equilibrium \cite{PhysRevE.107.024701} for the corresponding SRS aspect ratios. In particular, nematic ordering was observed for SRSs with aspect ratio \(L/D = 3\) and smectic ordering for \(L/D = 2\), despite the absence of such phases in equilibrium according to the Onsager criterion \cite{https://doi.org/10.1111/j.1749-6632.1949.tb27296.x}.

\textbf{d.} The influence of two-temperature activity becomes even more striking in systems of chiral rod-like particles \cite{chattopadhyay2024stability}, where the interplay between scalar activity and chirality introduces an additional dimension to the nonequilibrium phase behavior. In equilibrium, the formation of liquid-crystal phases such as nematic (N), cholesteric (\(N_c\)), smectic (Sm), and crystal (K) requires confining walls to induce planar alignment. Under two-temperature activity, however, the hot--cold interface itself acts as an effective aligning boundary, stabilizing liquid-crystal ordering in the cold domains even in the absence of external walls. Another particularly important observation concerns the destabilization and eventual breakdown of the cholesteric phase under scalar activity.


\section{Effect of confinement on 2-TIPS}

    \begin{figure*} 
    \onefigure[width = 0.85 \linewidth,height=9.5 cm]{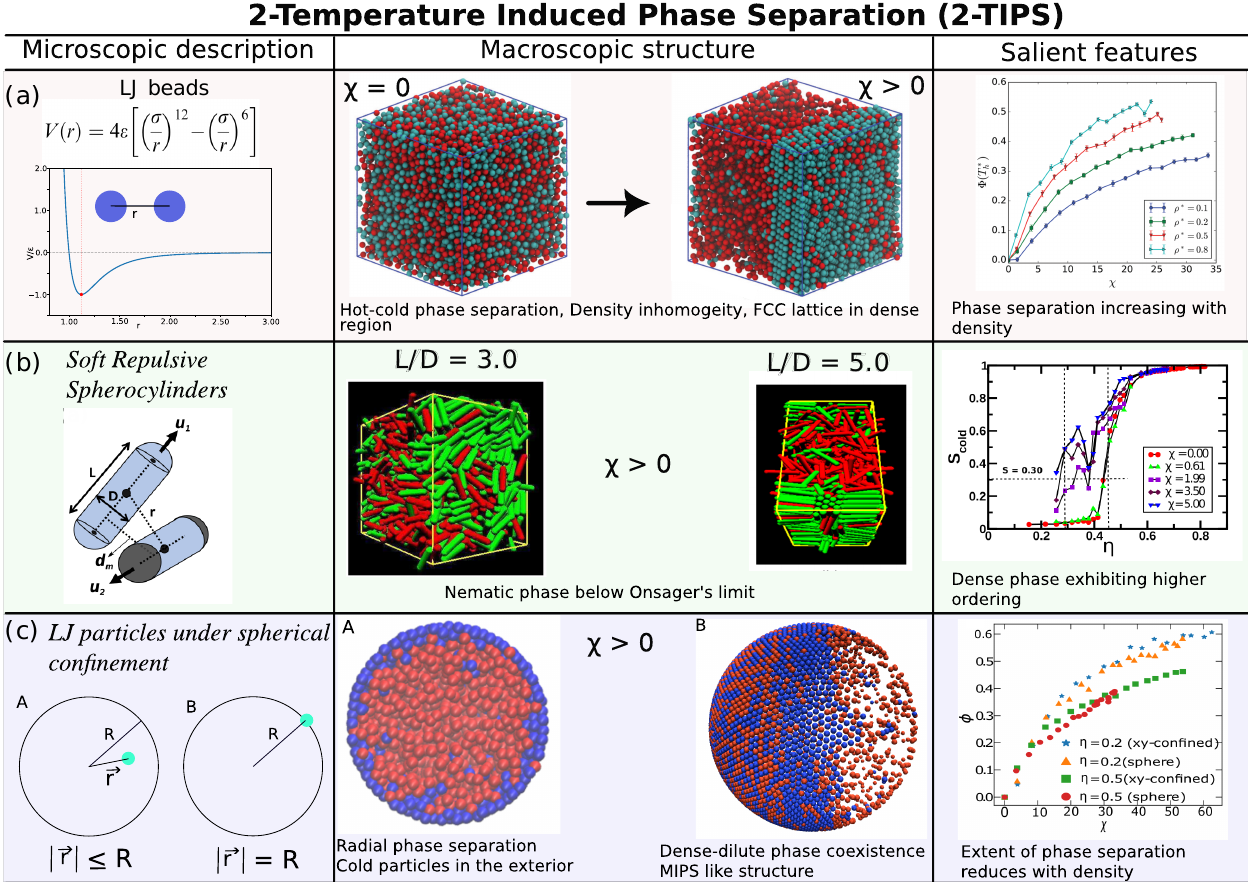}
    \caption{Details of the steady state structures in 2-TIPS for various soft matter systems. Panel (a) demonstrates the phase separation under two-temperature activity in a system of LJ particles in 3D \cite{chari2019scalar}. When shape anisotropy is introduced in the system, the non-equilibrium steady state structures stabilize the phases inaccessible in equilibrium \cite{chattopadhyay2021heating,PhysRevE.107.024701}, as shown in panel (b). Panel (c) demonstrates the effect of topological confinement on the 2-TIPS scenarios for LJ particles \cite{D3SM00796K}.}
    \label{fig:1}
    \end{figure*}

\begin{figure*}[ht!]
\centering
\includegraphics[width=0.73\textwidth,height=3.5cm]{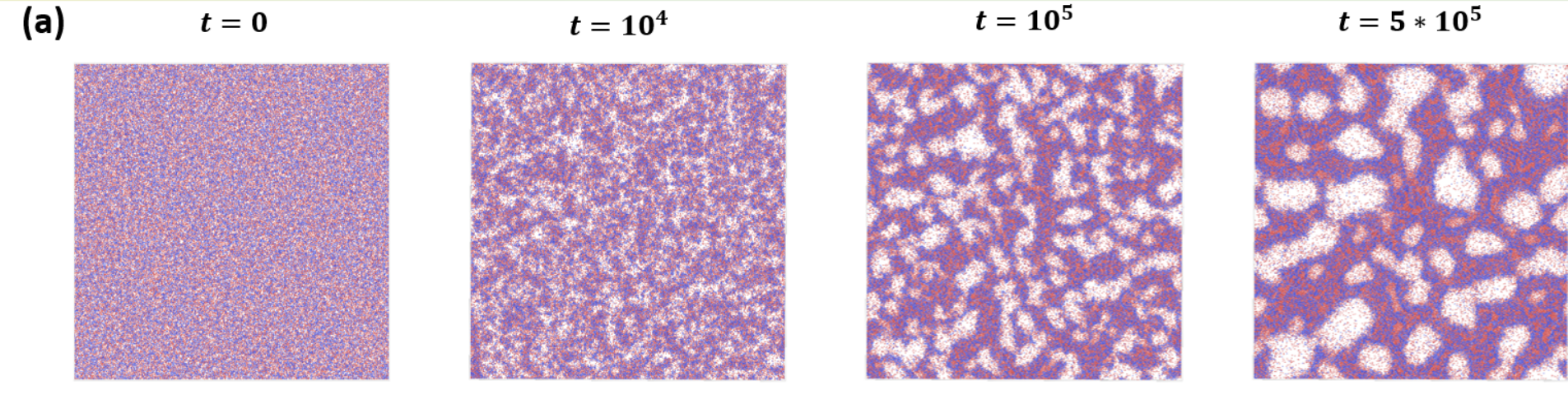}
\hfill
\includegraphics[width=0.24\textwidth,height=3.5cm]{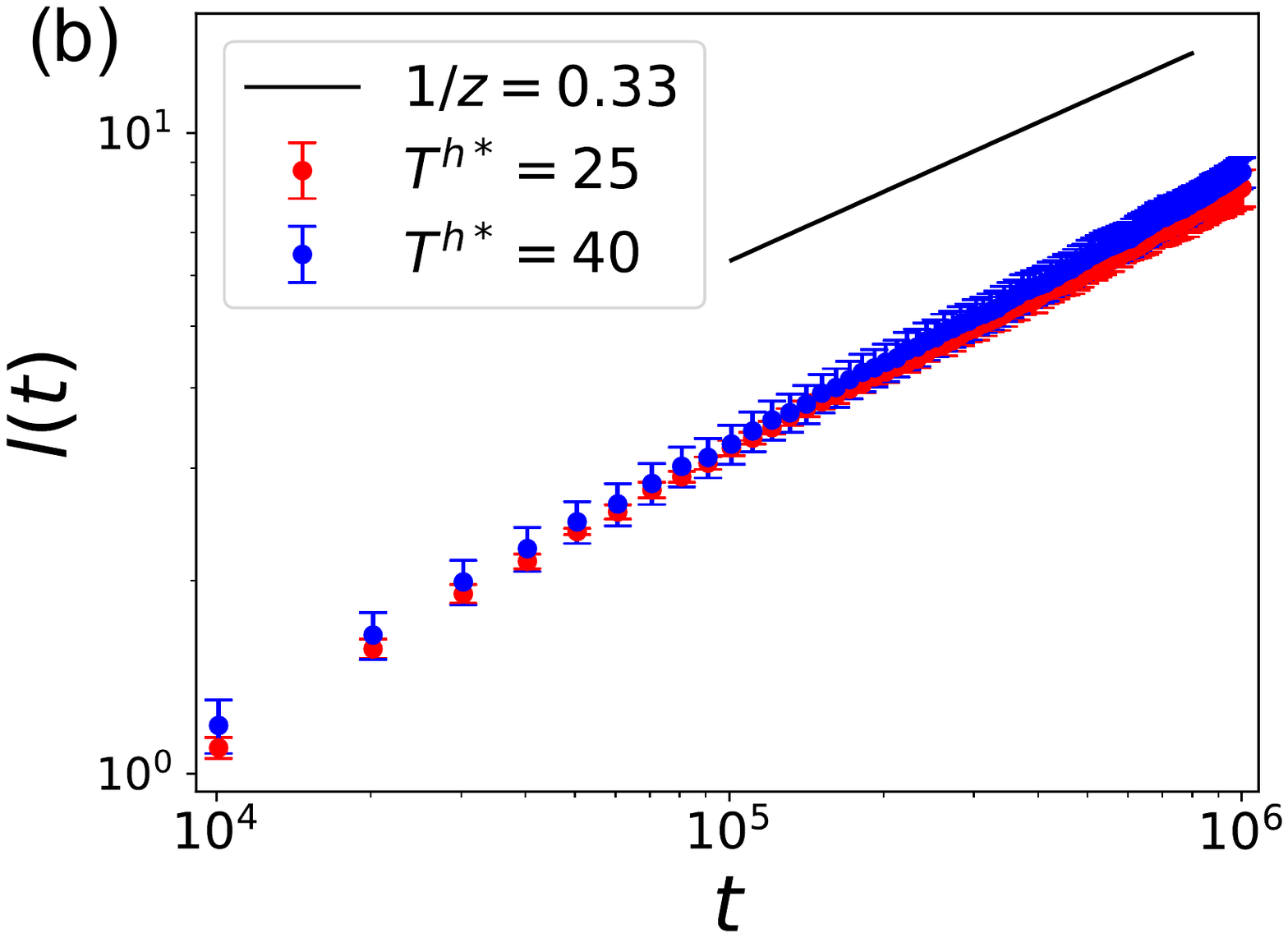}
\caption{Phase separation and coarsening in the high-density regime of 2-TIPS. (a) Snapshots showing the evolution of bicontinuous hot-rich (red) and cold-rich (blue) domains during phase separation. (b) Corresponding growth of the characteristic domain size, which follows a power law growth \(l(t)=t^{1/z}\), yields a growth exponent of \(1/z=0.33\) \cite{1c8b-hmxv}. (Adapted with permission from the American Physical Society publishing group.)}
\label{fig:kinetics}
\end{figure*}    
\textbf{a.} \textit{2-TIPS in the presence of walls:}  Similar to equilibrium systems \cite{mandal2025freezing,mandal2025melting,mandal2026phase,rajendra2023packing}, confinement significantly influences both the morphology and extent of phase separation in 2-TIPS \cite{D3SM00796K}, as illustrated in Fig. \ref{fig:1}(c). A key finding is that, despite the walls interacting identically and repulsively with both hot and cold particles, cold particles preferentially accumulate near the boundaries, as if subjected to an effective attraction toward the wall. This emergent wall affinity drives the formation of cold-rich clusters adjacent to parallel walls in both 2D and 3D. In contrast to bulk systems \cite{chari2019scalar}, under parallel confinement, the degree of phase separation decreases with increasing density. This behavior can be understood from the steady-state cluster composition: a substantial fraction of hot particles remains trapped within the dense cold-rich domains, and this fraction increases with density. The resulting increase in virial pressure hinders the migration of hot particles to the dilute phase, reducing the compositional contrast between dense and dilute regions and thereby weakening phase separation. 
  Under isotropic spherical confinement, however, the geometry induces radial segregation, with hot particles occupying the interior and cold particles accumulating near the boundary. In this case, the extent of phase separation increases with density and is significantly enhanced relative to the bulk. The curved confinement geometry provides a larger interfacial area between the hot- and cold-rich regions, allowing greater compositional segregation in the steady state. This highlights the important role of confinement geometry in determining the steady-state organization of two-temperature mixtures.
     
\textbf{b.} \textit{ 2-TIPS on constrained spherical surface:} For hot and cold particles constrained on spherical surface \cite{D3SM00796K}, the steady state structures and density effects are quite similar to the wall-confined planar systems. As the phase separation between the hot and cold particles decreases with density, the high-density steady state is characterised by a coexistence between dense and dilute regions - similar to MIPS, observed in Active Brownian Particles (ABPs) \cite{fily2012athermal,redner2013structure}. The 2-TIPS steady state structure also shows a local density-dependent velocity function, 
    like what is observed in MIPS. Despite the similarities, these two phenomena are fundamentally different in their origin; 2-TIPS incorporates two different velocity distributions for the two subparts in the system, whereas MIPS is observed in a system where all the particles have the same self-propelled velocities.

\subsection{Phase-separation kinetics in 2-TIPS}
The kinetics of phase separation \cite{bray1994theory,puri2009kinetics,hohenberg1977theory,PhysRevLett.96.016107,Majumder_2011} describes how an initially homogeneous system evolves into a phase-separated state following a quench into the phase-separated regime. In equilibrium binary mixtures, the studies on kinetics have revealed universal features such as dynamic scaling and universal growth laws that become independent of microscopic details at late times. An important question is whether such universality persists in intrinsically non-equilibrium systems undergoing 2-TIPS, where continuous energy transfer between hot and cold particles violates detailed balance. As discussed below, the kinetics of 2-TIPS exhibits rich density-dependent behavior \cite{1c8b-hmxv,venkatareddy2025phaseseparationkinetics2tips}, both in domain morphology and growth laws.

 Figure \ref{fig:kinetics} (a) illustrates the temporal evolution of domain morphologies in the binary mixture of hot and cold particles undergoing 2-TIPS in 2D. At high densities \cite{1c8b-hmxv}, the kinetics of 2-TIPS closely resembles spinodal decomposition in equilibrium binary mixtures. Following the quench, phase separation begins immediately through the formation of interconnected bicontinuous domains rich in either hot or cold particles. These domains subsequently coarsen with time while preserving their statistical self-similarity, as evidenced by the dynamic scaling of the spatial correlation function. The simulations reveal algebraic domain growth characterized by the Lifshitz–Slyozov growth exponent of \(1/3\) in both 3D and 2D, as depicted in Fig. \ref{fig:kinetics} (b). These observations demonstrate that, despite the intrinsically non-equilibrium nature of 2-TIPS, the coarsening kinetics at high density belongs to the same universality class as diffusive phase separation in passive binary mixtures. In contrast, the low-density regime \cite{venkatareddy2025phaseseparationkinetics2tips} exhibits a fundamentally different coarsening mechanism. Instead of bicontinuous domains, phase separation proceeds through the nucleation of isolated dense clusters of cold particles within a dilute hot background. These clusters migrate and coalesce, forming elongated, fractal-like domains. As a result, coarsening is significantly faster than in a high-density regime, with a growth exponent close to 0.7, well above the Lifshitz–Slyozov value. The enhanced growth arises from ballistic cluster agglomeration, which dominates the coarsening kinetics.

Interestingly, while the morphology of phase-separating domains in equilibrium binary mixtures is largely controlled by composition \cite{puri2009kinetics}, in 2-TIPS, the overall density plays an analogous role, determining whether the system forms bicontinuous domains or isolated clusters despite the equal concentration of hot and cold particles. It is also important to note that phase separation kinetics sensitively depends on the choice of thermostat. Weber \textit{et al.} \cite{weber2016binary} predicted a \(1/4\) growth law in a binary mixture governed by overdamped Brownian dynamics in 2D.

To obtain a mesoscopic understanding of the kinetics, phenomenological coarse-grained descriptions \cite{1c8b-hmxv,venkatareddy2025phaseseparationkinetics2tips} have also been developed for 2-TIPS. At high densities \cite{1c8b-hmxv}, a minimal model consisting of coupled conserved order-parameter equations for hot and cold particles successfully reproduces the bicontinuous domain morphologies, dynamic scaling, and Lifshitz–Slyozov growth law observed in MD simulations. In the low-density regime, the theory is extended by introducing a velocity field driven by an active stress (arising from the temperature gradient across the interphase), which captures ballistic motion and the coalescence of cold clusters, supporting ballistic agglomeration as the dominant growth mechanism.

\section{Conclusions and Outlook}

2-TIPS has emerged as a generic mechanism of non-equilibrium self-organization in systems composed of particles maintained at different temperatures. Unlike equilibrium phase separation or MIPS, 2-TIPS is driven solely by asymmetric energy injection and the resulting heat flux between particle species. Despite this simple microscopic origin, it gives rise to a rich range of collective phenomena, including spontaneous segregation into dense cold and dilute hot domains, activity-induced ordering, and nontrivial coarsening dynamics.

The studies reviewed here demonstrate that several features of 2-TIPS are universal across a broad class of soft matter systems. These include sustained entropy production, effective temperatures that differ from the imposed thermostat values, mechanical equilibrium through pressure balance, and the existence of a density-dependent critical activity for phase separation. At the same time, particle anisotropy, chirality, and confinement enrich the resulting phase behavior, leading to crystalline, liquid-crystalline, and chiral ordered states, as well as shifts of equilibrium phase boundaries and modified coexistence structures. Furthermore, the kinetics of 2-TIPS exhibits distinct density-dependent regimes, ranging from Lifshitz--Slyozov diffusive coarsening at high densities to ballistic cluster agglomeration at low densities.

Despite the considerable progress made in recent years, many important questions remain open. A fundamental theoretical challenge is the development of a unified coarse-grained description capable of connecting microscopic heat transfer, entropy production, and effective interactions to the macroscopic phase behavior observed across different systems. The role of thermostatting schemes and their influence on non-equilibrium steady states and coarsening dynamics also deserves further investigation. In addition, the origin of local temperature variations, pressure anisotropies, and the relationship between effective temperatures and thermodynamic observables remain poorly understood.

Given the diverse and strongly density-dependent coarsening behavior observed in 2-TIPS, an important future direction will be to investigate phase-separation kinetics in anisotropic systems such as mixtures of hot and cold SRSs. Such studies could reveal how orientational degrees of freedom and LC ordering influence domain morphology, growth laws, and coarsening mechanisms. More broadly, the interplay of scalar activity with confinement and topology and other non-equilibrium driving forces, including self-propulsion, alignment interactions, chemical gradients, and non-reciprocal interactions, presents a promising avenue for future research. On the experimental side, advances in colloidal, granular, and biological systems may provide opportunities to directly test theoretical predictions and establish 2-TIPS as a broadly relevant paradigm for non-equilibrium phase transitions.

Taken together, the results reviewed here suggest that 2-TIPS represents a robust and versatile route to non-equilibrium self-organization. By linking heat transfer, phase separation, and self-assembly within a common framework, it provides new insights into the collective behavior of scalar active matter and opens promising avenues for controlling structure formation far from equilibrium.

\section{Acknowledgements}
P.K.M thanks ANRF, India, for support through the JC Bose Grant (ANRF/JBG/2025/000227/PS).

\bibliographystyle{eplbib} 
\bibliography{review}
\end{document}